\documentclass[aps,prl,fleqn,floatfix,twocolumn,superscriptaddress]{revtex4}

\usepackage{graphicx}
\usepackage{amsmath,amssymb,amsfonts}

\begin{document}

\title{Stress Tensor and Bulk Viscosity in Relativistic Nuclear Collisions }

\author{Rainer J. Fries}
\affiliation{Cyclotron Institute and Department of Physics, Texas A{\&}M University, College Station, TX 77843, USA}
\affiliation{RIKEN/BNL Research Center, Brookhaven National Laboratory, Upton, NY 11973, USA}
\affiliation{Yukawa Institute of Theoretical Physics, Kyoto University, Kyoto 606-8502, Japan}
\author{Berndt M\"uller}
\affiliation{Yukawa Institute of Theoretical Physics, Kyoto University, Kyoto 606-8502, Japan}
\affiliation{Department of Physics, Duke University, Durham, NC 27708, USA}
\author{Andreas Sch\"afer}
\affiliation{Yukawa Institute of Theoretical Physics, Kyoto University, Kyoto 606-8502, Japan}
\affiliation{Institut f\"ur Theoretische Physik, Universit\"at Regensburg, D-93040 Regensburg, Germany}

\date{\today}

\begin{abstract}
We discuss the influence of different initial conditions for the stress tensor
and the effect of bulk viscosity on the expansion and cooling of the fireball 
created in relativistic heavy-ion collisions. In particular, we explore the 
evolution of longitudinal and transverse components of the pressure and 
the extent of dissipative entropy production in the one-dimensional, 
boost-invariant hydrodynamic model. We find that a bulk viscosity consistent
with recent estimates from lattice QCD further slows the equilibration of the
system, however it does not significantly increase the entropy produced.
\end{abstract}

\maketitle

\section{Introduction}

The formation of strongly interacting matter at supranuclear energy densities 
has recently been studied in great detail in nuclear reactions at the 
Relativistic Heavy Ion Collider (RHIC). The analysis of the data collected 
in these experiments
\cite{Arsene:2004fa,Back:2004je,Adams:2005dq,Adcox:2004mh}
has shown that the matter flows very rapidly at the moment of its breakup 
into free-streaming hadrons. The collective flow also exhibits a large 
anisotropy in non-central collisions, characterized by the ``elliptic'' flow 
parameter $v_2$. These observations are commonly understood to imply 
that (i) the quark and gluon matter undergoes rapid equilibration with 
thermalization times smaller than 1 fm/$c$ \cite{Heinz:2001xi}, and (ii) 
the quark gluon plasma (QGP) maintains low shear viscosity $\eta$ not 
much larger than the conjectured lower bound  
$4\pi\eta_{\rm KSS} = s$ where $s$ is the entropy density \cite{Kovtun:2004de}.
This has led to the claim that the quark gluon plasma formed at RHIC is 
the most perfect liquid known in nature \cite{Hirano:2005wx}.

The emerging picture still has some uncertainties. The predictions
of hydrodynamic simulations, especially for the flow anisotropy in off-central
collisions, depend on the assumed transverse density profile 
\cite{Hirano:2005xf}. Moreover,
the possible role of contributions to transverse flow from the pre-equilibrium 
phase of the reaction is not settled. Most hydrodynamic calculations assume
that no transverse flow is present at the time of initialization, usually chosen 
in the range $\tau = 0.5-1.0$ fm/$c$ after the onset of the reaction,
although there are good reasons to believe that transverse pressure gradients 
existing at earlier times will contribute to the generation of collective flow even 
if the parton momentum distribution is still anisotropic \cite{Fries:2007iy}.

Recently, it was pointed out that further complications could come from the 
bulk viscosity $\zeta$ of quark and gluon matter near the QCD phase 
transition \cite{Kharzeev:2007wb,Karsch:2007jc}. Bulk viscosity can be 
neglected compared to shear viscosity in many systems in nature. This was 
also shown to be true in 
quantum chromodynamics (QCD) at high temperature and weak coupling 
where $\zeta \sim \alpha_s^2 T^3/\ln \alpha_s^{-1}$ \cite{Arnold:2006fz}  
while $\eta \sim T^3/(\alpha_s^2 \ln \alpha_s^{-1})$ \cite{Arnold:2000dr}. 
The difference can be traced back to the near conformal invariance of 
QCD at high temperature. Near the pseudo-critical temperature $T_c$, 
however, QCD is far from being conformal, as can be inferred from the large 
peak of the ``interaction measure'' $(\epsilon -3P)/T^4$ at $T_c$ found 
in lattice QCD simulations \cite{Cheng:2007jq}. Indeed, the ratio of the bulk viscosity over
the entropy density, $\zeta/s$, on the lattice (in the quenched approximation) 
was recently found \cite{Meyer:2007dy}  to exhibit a narrow peak around $T_c$ 
of order unity with $\zeta/s \gg \eta/s$. The bulk viscosity can also be related 
directly to the trace anomaly of the energy momentum tensor, and an estimate 
can thereby be obtained from lattice results for the interaction measure 
\cite{Karsch:2007jc}. 

The large spike of the bulk viscosity near $T_c$ immediately raises several 
questions \cite{Karsch:2007jc} which we want to address here. However, we 
also want to take a look at the larger picture. Ideal hydrodynamics requires 
complete thermalization of the matter. Isotropization of the pressure $P$ 
and consistency with the equation of state $P(\epsilon)$ are 
necessary conditions for equilibrium. Viscous hydrodynamics permits 
certain kinds of small deviations from equilibrium. 
In the local rest frame, shear stress $\pi^{ij}$ describes deviations from 
isotropy of the stress tensor while bulk stress $\Pi\delta^{ij}$ measures 
deviations from the equilibrium equation of state. In the second-order 
formulation of viscous
hydrodynamics \cite{Muller:1967,Israel:1979wp}, the deviations of the
stress tensor from its equilibrium form are not prescribed by external strains,
such as flow shear or flow divergence, but can be independently given and
only relax to the externally forced values over time. The fireball produced in 
a relativistic nuclear collision starts out with highly anisotropic particle 
distributions and therefore a large pressure anisotropy, which may not be
related to the imprinted flow field. It is thus useful to explore how the 
hydrodynamical evolution responds to different initial deviations from 
local equilibrium.

It is not always obvious how large the deviations can be before the 
hydrodynamical approximation fails. In principle, viscous corrections extend 
the reach of hydrodynamics in relativistic heavy ion collisions to earlier times, 
but the gradient expansion of the stress tensor may break down 
\cite{Lublinsky:2007mm}.  Similarly, a large bulk viscosity $\zeta$ around 
$T_c$ could mean that the matter is driven far from equilibrium around the 
phase transition and may even develop dynamical instabilities 
\cite{Torrieri:2008ip}. It has also been claimed \cite{Karsch:2007jc} that bulk 
stress could contribute significantly to entropy production, with possibly 
profound consequences for the hadronization mechanism at $T_c$.
Furthermore, one expects the viscous bulk pressure $\Pi$ to be negative for 
an expanding system ($\Pi = -\zeta\partial_\mu u^\mu$ to first order in
gradients), implying that
the effect of bulk viscosity is to slow down the expansion of the system.
This should be most relevant in the longitudinal direction where
the pressure is also reduced by the viscous shear stress.  Hence, 
one expects the system to spend more time around $T_c$
than predicted by ideal hydrodynamics. This leads to yet another 
interesting prospect: Perhaps the evolution of the system is more sensitive
to the equation of state near $T_c$ and the order of the phase transition than 
one would expect from ideal hydrodynamics where the relation of the phase
transition to observables has been found to be rather obscure 
\cite{Rischke:1996em}.

Here, want to explore these questions in a hydrodynamic
model with a simple space-time structure, but realistic
equation of state, and bulk and shear viscosities motivated by lattice QCD.
Our framework is a one-dimensional boost-invariant fireball 
with translational and rotational symmetries in the transverse plane
as first described by Bjorken \cite{Bjorken:1982qr}.  We discuss the 
time evolution of the components of the pressure using second order
hydrodynamics. Our treatment is comparable to that of Baym 
\cite{Baym:1984np} and Heiselberg and Wang 
\cite{Heiselberg:1995sh,Heiselberg:1996xg},
who explored deviations from, and the approach to, equilibrium in a
boost-invariant expansion in the framework of the relaxation time
approximation to the Boltzmann equation. We will also explore the
consequences of several different assumptions about the initial longitudinal 
and transverse pressure.

Let us rephrase our catalog of questions in the context of our hydrodynamic
model:
(i) How far from the equilibrium pressure $P$ is the longitudinal 
pressure $P_z$ throughout the evolution and in particular at the phase 
transition?
(ii) How close to isotropy is the pressure tensor, i.e. how large is $|P_{\perp}-P_z|/P$?
(iii) What is the amount of entropy $S_\Pi$ produced by the bulk viscosity 
compared to contributions from shear viscosity $S_\Phi$ during the lifetime 
of the fireball? 

We are aware that the simplified treatment with 1-dimensional
expansion has several limitations. This approximation does not permit us 
to explore effects related to the transverse expansion and the more
rapid cooling coming from it. We also cannot treat the flow anisotropy in
the transverse plane seen in off-central collisions. Since boost-invariant
hydrodynamics tends to overestimate the time required for the 
matter to cool below $T_c$ it makes our study conservative in the 
sense that the effects of the bulk viscosity are underestimated, compared 
with those expected in a full three-dimensional treatment.

\section{Viscous Hydrodynamics}

In the center of the fireball in a nuclear collision the viscous 
stress-energy tensor in the local comoving frame has the form 
\cite{Muronga:2001zk,Teaney:2003kp,Muronga:2004sf}:
\begin{equation}
T^{\mu\nu} = \left(
\begin{array}{cccc}
\varepsilon & 0 & 0 & 0 \\
0 & P_{\perp} & 0 & 0 \\
0 & 0 & P_{\perp} & 0 \\
0 & 0 & 0 & P_{z} 
\end{array} \right) 
\label{eq:Tmunu}
\end{equation}
with the transverse and longitudinal pressure
\begin{eqnarray}
P_{\perp} &=& P + \Pi + \frac{1}{2}\Phi 
\nonumber \\
P_{z} &=& P + \Pi - \Phi
\label{eq:stress}
\end{eqnarray}
Here $P$ denotes the (isotropic) pressure in thermal equilibrium, 
$\Phi$ and $\Pi$ denote the non-equilibrium contributions to the pressure
coming from shear and bulk stress. In particular, the traceless
shear tensor in that frame takes the form 
$\pi^{ij} = \mathrm{diag}(\Phi/2, \Phi/2,-\Phi)$
consistent with the symmetries in the transverse directions.
We refer the reader to Refs.~\cite{Muronga:2001zk,Heinz:2005zi,Muronga:2004sf}
for more details. 

At early times, $\Phi$ and $\Pi$ will be given by the initial conditions for 
the stress tensor established by the mechanisms of energy and momentum 
deposition in the nuclear collision.  Reflecting the dilution effect of the 
expansion on the local longitudinal momentum distribution of partons, 
the pressure components are expected to satisfy the ordering 
\begin{equation}
P_{z} \equiv T^{zz} < T^{xx} = T^{yy} \equiv P_{\perp}.
\label{eq:pressure}
\end{equation} 
One can argue that the largest physically meaningful value of $\Phi$ at a 
time $\tau$ is $\Phi = 4(P+\Pi)$, which corresponds to $P_{z} = - P_{\perp}$. 
Such a negative value of the longitudinal pressure arises, for example, 
when the matter is completely in the form of coherent longitudinal fields 
at very early times \cite{Fries:2007iy,Kovchegov:2007pq}. However, such
a configuration is very far from equilibrium and the hydrodynamic
approximation is surely invalid. Once decoherence of the field is
reached at a time $\tau_{\mathrm{dec}}$ the components of the physical 
pressure should be positive: $P_z, P_\perp \geq 0$.
We want to explore various scenarios for the time evolution of $\Phi$ and 
$\Pi$, assuming initial values satisfying $\Phi \leq P+\Pi$ at the start
time of the hydrodynamic evolution $\tau_0 \ge \tau_{\mathrm{dec}}$. 

In first-order (Navier-Stokes) dissipative hydrodynamics the bulk and shear 
stress are related to gradients in the system via the bulk and shear 
viscosities through
\begin{equation}
  \label{eq:NS}
  \Pi = -\zeta \partial^\mu u_\mu \, , \quad \pi^{\mu\nu} = 2\eta 
   \nabla^{\langle\mu} u^{\nu\rangle} \, ,
\end{equation}
where $u^\mu = (\cosh \eta, 0 ,0 \sinh\eta)$ in our case is the expansion 
velocity, $\eta$ is the space-time rapidity, $\nabla^\mu = \partial^\mu -
u^\mu (u_\nu \partial^\nu)$, and $\langle \ldots \rangle$ indicates a
projection orthogonal to $u^\mu$, symmetrization of indices and removal 
of the trace.

We follow the spirit of the (Israel-Stewart) theory of second-order 
dissipative hydrodynamics \cite{Israel:1979wp} by assuming that the 
actual bulk and shear stress have the freedom to relax to their
first-order values at rates governed by relaxation times $\tau_\Pi$
and $\tau_\pi$.
The equations governing the longitudinal expansion of the medium in
our case are then 
given by \cite{Heinz:2005zi,Muronga:2003ta,Baier:2006um,Baier:2007ix}:
\begin{eqnarray}
  \frac{\partial\varepsilon}{\partial\tau} &=& - \frac{1}{\tau}(\varepsilon 
  + P + \Pi - \Phi) \, ,  
  \label{eq:evol1} \\
  \tau_{\pi} \frac{\partial\Phi}{\partial\tau} &=& \frac{4\eta}{3\tau} -  \Phi(\tau) -
  \left[ \frac{4\tau_{\pi}}{3\tau}\Phi +\frac{\lambda_1}{2\eta^2}\Phi^2\right] 
  \, , 
  \label{eq:evol2} \\
  \tau_{\Pi} \frac{\partial\Pi}{\partial\tau} &=& - \frac{\zeta}{\tau} - 
  \Pi(\tau) .
\label{eq:evol3}
\end{eqnarray}
It was recently pointed out by Baier et al.\ \cite{Baier:2007ix} that the terms 
in the square bracket in (\ref{eq:evol2}) are required in a theory with 
conformal symmetry. Conformal symmetry is approximately realized in 
QCD at high temperatures. Since these terms have not been studied
quantitatively we will examine their influence on the evolution below.

The entropy density $s$ obeys the equation \cite{Heinz:2005bw}:
\begin{equation}
  \frac{\partial(\tau s)}{\partial\tau} = \frac{\tau}{T} \left(
  \frac{3 \Phi^2}{4 \eta} + \frac{\Pi^2}{\zeta} \right).
  \label{eq:dsdtau}
\end{equation}
$\tau s = dS/(dydA)$ is the entropy per unit rapidity $y$ and transverse area
$A$. $\tau s$ is constant for ideal hydrodynamics.
Bulk and shear stress relax toward their Navier-Stokes
values. The late time behavior of these values for 1-dimensional
boost-invariant expansion is then directly given by Eq.\ (\ref{eq:NS}) as
\begin{equation}
\Phi = \frac{4\eta}{3\tau} ,\qquad \Pi = - \frac{\zeta}{\tau} \, .
\label{eq:eta-zeta}
\end{equation}
The larger the shear viscosity, the more anisotropic the pressure remains
at late times.  

Relaxation times and viscosities are related by coefficients $\beta_0$
and $\beta_2$ which are determined by the underlying theory:
\begin{equation}
  \tau_\Pi = \zeta \beta_0 \, , \quad \tau_\pi = 2\eta \beta_2
\end{equation}
Kinetic theory of massless partons predicts a value, $\beta_2 = 3/(4P)$ 
\cite{Baier:2006um,Muronga:2003ta}. This leads to a relaxation time
which is roughly given by
\begin{equation}
  \tau_\pi^{\rm (kin)} = \frac{3}{2\pi T}
  \label{eq:taueq1}
\end{equation}
In conformal hydrodynamics a different behavior is obtained by
matching the asymptotic form of a boost-invariant, longitudinally 
expanding thermal medium in the $N=4$ supersymmetric 
Yang-Mills theory to a hydrodynamic evolution 
\cite{Baier:2007ix,Natsuume:2007tz}
\begin{equation}
  \tau_{\pi}^{\rm (SYM)} = \frac{2-\ln 2}{2\pi T} .
  \label{eq:taueq2}
\end{equation}
Obviously the value of the relaxation time from kinetic theory is about 
twice as long as the latter. We will test both values below.
In absence of further reliable predictions, we will always assume the same 
relaxation time for the bulk stress as a function of temperature, 
$\tau_\Pi (T) = \tau_\pi (T)$.
The coefficient $\lambda_1$ in (\ref{eq:evol2}) was determined 
for supersymmetric Yang-Mills theory to be
\cite{Baier:2007ix}
\begin{equation}
  \lambda_1 = \frac{\eta}{2\pi T} \, .
\end{equation}

\section{Viscosities and Initial Conditions}

We have already specified the equations of motion and our choice for the
parameters $\tau_\pi$, $\tau_\Pi$ in the previous section. We further assume 
that the matter is characterized by a minimal shear viscosity, i.e.\ we set 
$\eta = \eta_\mathrm{KSS} = s/(4\pi)$. This is not in contradiction to 
lattice QCD 
results \cite{Meyer:2007ic}, which lie close to the KSS bound. The pressure 
anisotropies found in our calculation can therefore be considered as a 
\emph{lower bound}.  For the equilibrium equation of state we use  
recent lattice QCD results for two light and one heavy quark flavors 
\cite{Cheng:2007jq}. We have parameterized the reduced equilibrium 
pressure $P/T^4$ and interaction measure $(\varepsilon-3P)/T^4$ for 
$N_\tau =6$ lattices from this reference.  The critical temperature is 
$T_c = 196$ MeV.

For the bulk viscosity we explore several options. Our starting point is
a recent calculation by Meyer in quenched QCD \cite{Meyer:2007dy}.
This calculation is not directly compatible with our equation of state from 
unquenched QCD. To deal with this problem, we have here chosen to parameterize 
the dimensionless ratio $\zeta/s$ as a function of the dimensionless ratio 
$\omega = (\varepsilon-3P)/(\varepsilon+P)$. This choice corrects for the shift in
the critical temperature, and it softens the very steeply peaked behavior of
$\zeta/s$ found by Meyer in quenched QCD. We call the bulk viscosity 
resulting from this fit $\zeta_0$. We note that several mechanisms can 
contribute to the bulk viscosity \cite{Paech:2006st}. In QCD, $\zeta/s$ is not only 
a function of the interaction measure but also of the speed of sound, the quark
condensates, and the relaxation time scale of the compression mode
\cite{Karsch:2007jc,Jeon:1994if}.  We also note that the statistical and systematic 
uncertainties of the existing lattice results for quenched QCD are quite large. 
We account for these combined uncertainties by running the hydrodynamic 
evolution for several values of $\zeta$ which have been obtained from $\zeta_0$  
by multiplying with a scaling factor $c_\zeta$: $\zeta = c_\zeta \zeta_0$, and by
changing the width of the peak near $T_c$ by a scale factor $1/a_\zeta$.

\begin{figure}[tb]   
\centerline{
\includegraphics[width=0.95\linewidth]{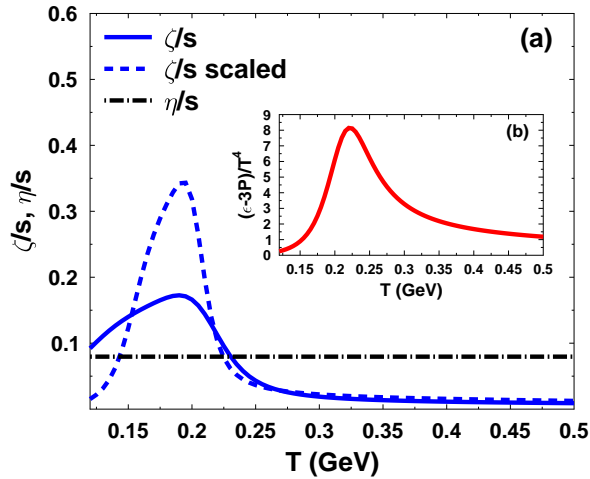}
}
\caption{(Color online) 
Kinematic bulk viscosity $\zeta/s$ and kinematic shear viscosity $\eta/s$ as a function 
of temperature $T$. $\eta/s$ (dash-dotted line, black) is determined by the KSS 
bound $\eta/s = 1/(4\pi)$, $\zeta/s$ (solid blue) is derived by a fit to the results 
reported in \cite{Meyer:2007dy}, using the conformal measure $\omega$ as
scaling variable. This fit is denoted as $c_\zeta=a_\zeta=1$. The blue dashed 
curve shows a modified parametrization for $\zeta/s$ with double the peak
height ($c_\zeta=2$) and half the peak width ($a_\zeta=2$).
Insert: parameterization of the reduced interaction measure 
$(\epsilon-3P)/T^4$ taken from \cite{Cheng:2007jq} as a function of $T$.}
\label{fig1}
\end{figure}

We show the results of our fits in Fig.\ \ref{fig1}. The kinematic bulk viscosity 
fitted from the lattice results as a function of temperature (shown by the solid 
(blue) curve) is compared with the kinematic shear viscosity from the KSS 
relation $\eta/s= 1/(4\pi)$ (dash-dotted (black) line). The dashed (blue) curve
shows a modified parametrization for $\zeta/s$ with double the peak height 
and half the peak width ($c_\zeta=a_\zeta=2$). The bulk viscosity exceeds 
the shear viscosity for temperatures below 220 MeV. The insert shows the 
interaction measure fitted from the results of Cheng {\em et al} \cite{Cheng:2007jq}.

For all simulations we choose a starting time $\tau_0 = 0.3$ fm/$c$ which is 
smaller than the equilibration times estimated from ideal hydrodynamics and 
is compatible with the expected decoherence time 
\cite{Muller:2005yu,Fries:2008xx} of the initial gluon field. 
The initial energy density is fixed to be
$\epsilon(\tau_0) = 50$ GeV/fm$^3$ which corresponds to an initial equilibrium 
temperature of roughly 400 MeV. We discuss three different initial 
conditions: 
\begin{itemize}
\item[(i)] ``equilibration'' sets $\Pi(\tau_0) = \Phi(\tau_0) = 0$.
\item[(ii)] ``1$^\mathrm{st}$ order'' uses the values given by first-order 
viscous hydrodynamics, $\Pi(\tau_0) = -\zeta(T_0)/\tau_0$,
$\Phi(\tau_0) = 4\eta(T_0)/(3\tau_0)$ where $T_0$ is the initial 
temperature at $\tau_0$.
\item[(iii)] ``anisotropic'' uses $\Pi(\tau_0) = -\zeta(T_0)/\tau_0$ as in 
(ii), but fixes $\Phi(\tau_0) = P(\tau_0)+\Pi(\tau_0)$ for vanishing 
initial longitudinal pressure.
\end{itemize}

We also remind the reader that we will run the hydrodynamic evolution both
with the conformal terms in Eq.\ (\ref{eq:evol2}) (we will denote this scenario
by the label ``C'') and in a standard version without them (denoted by ``S'').
We also explore both the short (``SYM'') and long (``kin'') relaxation time.
Each run below will be denoted by a 4-component label indicating the set of 
initial conditions, relaxation times, absence or presence of the conformal 
terms and the scaling variable $c_\zeta = \zeta/\zeta_0$.

\section{Results}

\begin{figure}[tb]   
\centerline{
\includegraphics[width=\columnwidth]{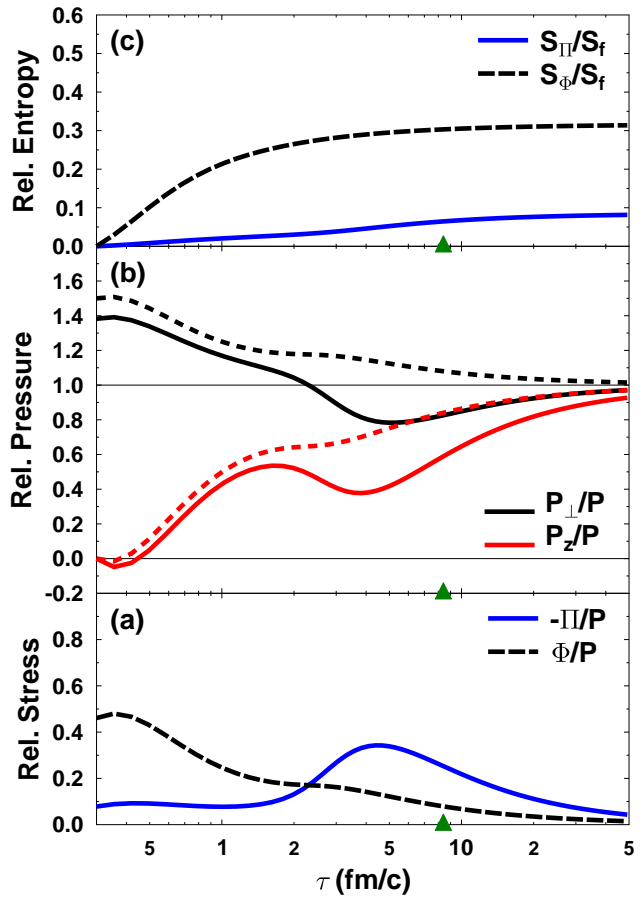}
}
\caption{(Color online)
(a) Relative bulk and shear stress, $-\Pi/P$ (solid blue)
and $\Phi/P$ (dashed black), as functions of time $\tau$ for scenario
(iii,S,SYM,$c_\zeta=1$), i.e. anisotropic initial conditions, standard
evolution without conformal terms and short relaxation times. The time 
when $T_c$ is reached is indicated by the triangle.
(b) Relative transverse and longitudinal pressure, $P_\perp/P$
(red) and $P_z/P$ (black), as functions of time $\tau$
for (iii,S,SYM,$c_\zeta=1$) (solid lines) and for the same scenario but with 
$c_\zeta$ set to zero (dashed lines).
(c) Relative entropy production from bulk and shear stress,
$S_\Pi/S_f$ (solid blue) and $S_\Phi/S_f$ (dashed black) as functions of 
time $\tau$ for the same scenarios with $c_\zeta=1$.
All results are for $a_\zeta=1$.}
\label{fig2}
\end{figure}

In this section we present the results of numerical solutions
of the viscous hydrodynamical equations (\ref{eq:evol1}--\ref{eq:evol3}).
We start by checking the influence of different values of the bulk viscosity 
by varying $c_\zeta$ with $a_\zeta$ fixed. We compare vanishing bulk 
viscosity ($c_\zeta=0$) with the lattice inspired value ($c_\zeta=1$).
Fore these runs we fix the set of initial conditions to anisotropic 
(iii), use evolution without the conformal terms (S), and choose the short 
relaxation time (SYM).
Figure \ref{fig2} shows the development of different components of the
pressure relative to the equilibrium pressure $P$ as a function of time. 
Part (a) shows the dissipative stress components $-\Pi/P$ and 
$\Phi/P$ due to bulk and shear viscosity, respectively, for $c_\zeta=1$. 
As expected the effects of shear viscosity are dominant at 
early times while bulk stress peaks later, when $\zeta/s$ starts to rise
around $T_c$. The peak in $-\Pi/P$ is reached already slightly
before the critical temperature $T_c$ --- the time $\tau_c \approx 8.4$ fm/$c$
when the system reaches $T_c$ (for $c_\zeta=1$) is indicated by the solid 
triangles on the abscissae.

Figure \ref{fig2} (b) displays the relative longitudinal and transverse pressure
$P_\perp/P$ and $P_z/P$, respectively for $c_\zeta=1$ (solid lines) and 
$c_\zeta=0$ (dashed lines). Both quantities develop a pronounced 
minimum just above $T_c$ if bulk viscosity is present, which corresponds to 
the maximum in $-\Pi/P$. The minima above $T_c$ are absent for vanishing 
bulk viscosity. We observe that the system is still very anisotropic even 
at $T_c$, and with $c_\zeta =1$ both pressure components are well below the 
equilibrium value. 

As mentioned earlier, the stability of boost invariant relativistic
hydrodynamics in the presence of a bulk viscosity was recently studied
by Torrieri and Mishustin \cite{Torrieri:2008ip}. Their treatment differs from
ours in two important details: They study the first-order (Navier-Stokes)
formulation of viscous hydrodynamics, and they use a parametrization of the
temperature dependence of the bulk viscosity which is much more strongly
peaked near $T_c$ and attains a much higher peak value. Even for their 
most conservative choice ($z_0=0.1$ in the notation of \cite{Torrieri:2008ip}), 
the peak in $\zeta/s$ is about twenty times higher than our peak value and 
five times as high as the largest value obtained by Meyer \cite{Meyer:2007ic}. 
We have solved our set of equations 
with their parameters and used a very small relaxation time $\tau_\Pi = \tau_\pi$ 
to emulate the Navier-Stokes limit. We find that $P_z$ becomes strongly negative 
in the range where $\zeta/s$ peaks, indicating that the matter is not only 
hydrodynamically unstable, as found in ref.\ \cite{Torrieri:2008ip}, but also 
thermodynamically unstable. This behavior is obviously a result of the highly 
peaked parametrization adopted by the authors of ref.\ \cite{Torrieri:2008ip}.

We now return to our own study. Figure \ref{fig2} (c) 
shows the entropy per unit rapidity and transverse area
produced by shear and bulk viscous effects, $S_\Phi = \tau s_\Phi$ and 
$S_\pi = \tau s_\Pi$, respectively. They correspond to the first and second 
term in Eq.\ (\ref{eq:dsdtau}) and are shown relative to the total final 
value $S_f = \tau_f s(\tau_f)$ where the final time $\tau_f$ 
is fixed at 50 fm/$c$. We note that despite its dramatic effects on the 
longitudinal pressure the contribution of the bulk stress to entropy 
production is rather moderate. The entropy produced by shear stress is much 
larger, due to the large velocity gradient in the initial state which 
generates large dissipative effects. However, the majority of the entropy 
production is confined to the earliest time period $\tau_0<\tau<1$ fm/c, 
suggesting that a hydrodynamical description of the matter rapidly loses 
reliability before 1 fm/c. We also remind the reader that our results are 
obtained for the minimal value of the shear viscosity ($\eta = s/(4\pi)$). 
Had we chosen a larger shear viscosity, the produced entropy would be 
respectively larger.

\begin{figure}[tb]   
\centerline{
\includegraphics[width=\columnwidth]{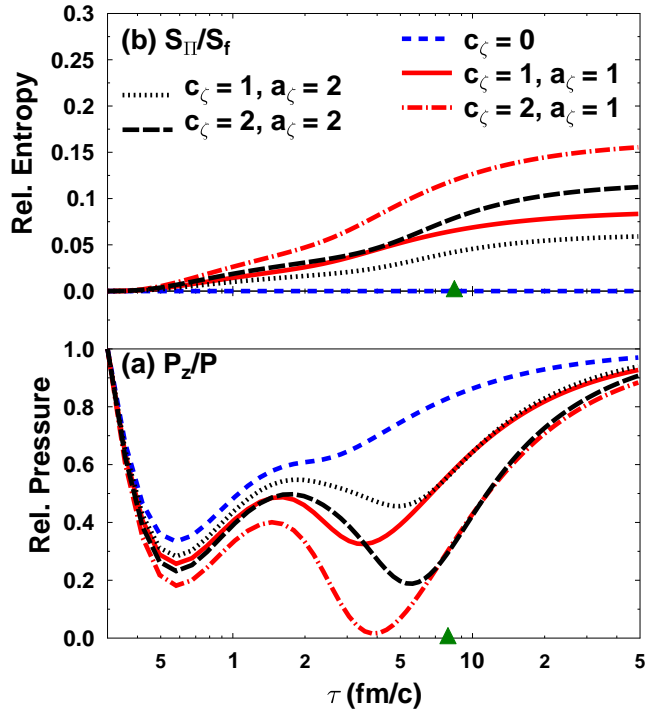}
}
\caption{(Color online)
(a) Relative longitudinal pressure $P_z/P$ as function of time $\tau$ for 
scenario (i,S,SYM) and various choices for the bulk viscosity as shown
in the legend ($c_\zeta=0,1,2$ and $a_\zeta=1,2$. The solid line shows
our ``standard'' parametrization ($c_\zeta=a_\zeta=1$).
(b) Relative entropy production from bulk stress, $S_\Pi/S_f$ as function 
of time $\tau$ for the same scenarios as in part (a).}
\label{fig2a}
\end{figure}

In Fig.\ \ref{fig2a} we have varied the parameters $c_\zeta$ and $a_\zeta$,
influencing the height and width of the peak in the bulk viscosity near $T_c$,
respectively, for the scenario (i,S,SYM). The short-dashed (blue) curve in
Fig.\ \ref{fig2a}(a) shows the relative longitudinal pressure in the absence 
of any bulk viscosity for comparison. The solid
and dash-dotted (red) lines are obtained for our parametrization of $\zeta(T)$ 
as shown by the solid (blue) line in Fig.\ \ref{fig1}, with its height scaled by
the factor $c_\zeta=1,2$. Obviously, the relative longitudinal pressure drops 
to almost zero above $T_c$ for $c_\zeta=2$ indicating that the local equilibrium 
assumption begins to break down in this case. The dotted and long-dashed
(black) curves correspond to the parametrization ($a_\zeta=2$) for a narrower 
peak in $\zeta(T)$, as shown by the dashed (blue) line in Fig.\ \ref{fig1}. In this 
case the onset of the strong reduction in $P_z/P$ is delayed compared with the 
case $a_\zeta=1$, but the effect in the immediate vicinity of $T_c$ (the location
is indicated by the (green) triangle on the abscissa) is found to be mainly sensitive
to the height of the peak, parametrized by $c_\zeta$, not its width. It is important
to note, however, that the hydrodynamical evolution becomes increasingly sensitive
to the precise value of the equilibration time for the bulk viscosity, $\tau_\Pi$, as
the width of the peak in $\zeta(T)$ becomes narrower. The small value chosen
here ($\tau_\Pi=\tau_\pi$) may be inappropriate for a very narrow peak. Insofar
as a narrow, high peak in $\zeta(T)$ is indicative of a near-critical behavior of
the medium near $T_c$, one would expect any mode that participates in this 
behavior to exhibit critical slowing down and its relaxation time to increase.

\begin{figure}[tb]   
\centerline{
\includegraphics[width=\columnwidth]{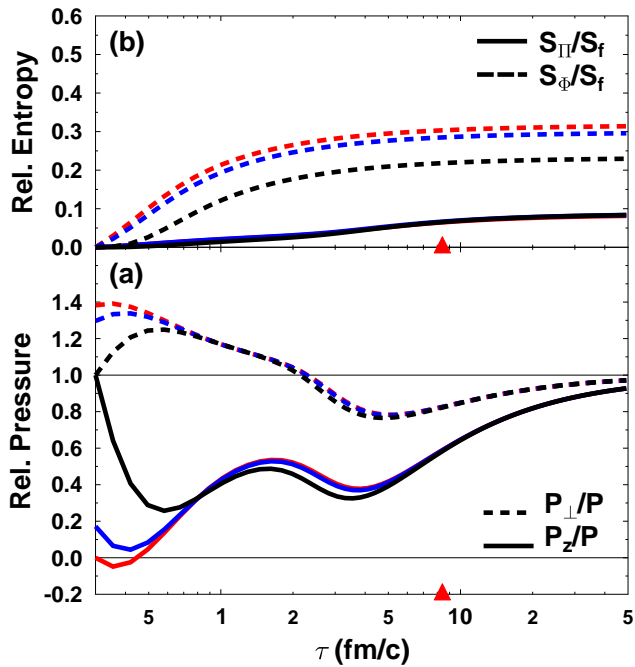}
}
\caption{(Color online)
(a) Relative transverse and longitudinal pressure, $P_\perp/P$ (dashed) 
and $P_z/P$ (solid), as functions of time $\tau$ for (S,SYM,$c_\zeta=1$) 
and initial conditions (i) (black), (ii) (blue), and (iii) (red). 
(b) Relative entropy 
production from bulk and shear stress, $S_\Pi/S_f$ (solid) and $S_\Phi/S_f$ 
(dashed) as functions of time $\tau$ for the same set of scenarios. 
Note that the different curves for $S_\Pi/S_f$ lie almost on top of each other.
The triangle indicates the time of critical temperature for initial condition
(iii). For conditions (i) and (ii) $T_c$ is reached slightly earlier.
All curves are for $a_\zeta=1$.}
\label{fig3}
\end{figure}

We now compare the impact of different initial conditions on the standard
evolution (S) for a bulk viscosity given by the scaling factors $c_\zeta=
a_\zeta=1$ and short relaxation time (SYM). Figure \ref{fig3} (a)
shows the relative pressure components $P_z/P$ and $P_\perp/P$. The 
most noticeable feature here is that even for  equilibrium initial conditions 
(i) the strong gradients in longitudinal 
direction drive the system immediately off equilibrium. The first
order initial conditions (ii) are very close to the maximal anisotropic
initial conditions (iii) which were loosely extrapolated from classical gluon 
fields. 
Interestingly, the effect of the different initial conditions is wiped out 
after a very short time $\Delta \tau \approx 0.5$ fm/$c$ and the system evolves 
in an universal way from that time forward. However, the different initial 
conditions for $\Phi$ leave a trace in the entropy produced during this 
stage of the evolution, as can be seen in Fig.\ \ref{fig3} (b).
On the other hand, the entropy production from bulk stress picks up most 
contributions around $T_c$ and is independent of the initial conditions.

\begin{figure}[tb]   
\centerline{
\includegraphics[width=\columnwidth]{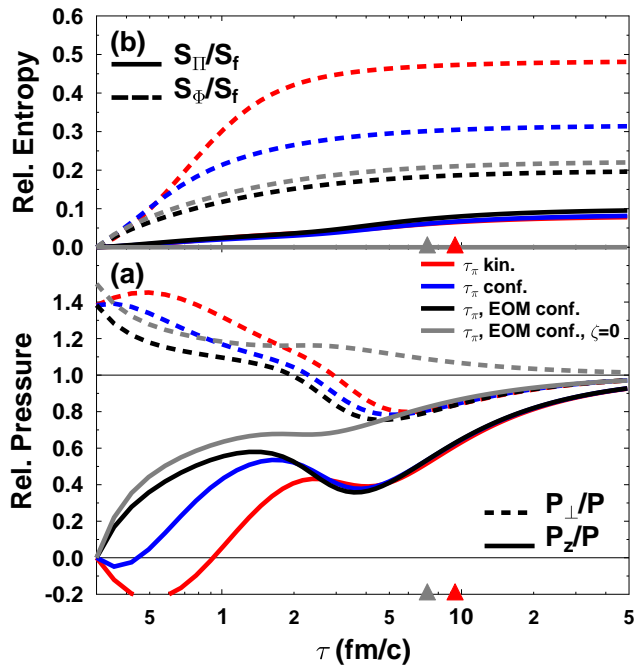}
}
\caption{(Color online)
(a) Relative transverse and longitudinal pressure, $P_\perp/P$ (dashed) 
and $P_z/P$ (solid). 
(b) Relative entropy 
production from bulk and shear stress, $S_\Pi/S_f$ (solid) and $S_\Phi/S_f$ 
(dashed). The different scenarios shown for initial conditions (iii) are:
$\tau_\pi=\tau_\Pi=\tau_\pi^{\rm (kin)}$ from kinetic theory, see Eq.~(\ref{eq:taueq1}), 
no conformal terms in (\ref{eq:evol2}), $c_\zeta=1$ [red]; 
$\tau_\pi=\tau_\Pi=\tau_\pi^{\rm (SYM)}$ from conformal symmetry, see Eq.~(\ref{eq:taueq2}), 
no conformal terms in (\ref{eq:evol2}), $c_\zeta=1$ [blue];
$\tau_\pi=\tau_\Pi=\tau_\pi^{\rm (SYM)}$, 
conformal terms in (\ref{eq:evol2}) switched on, $c_\zeta=1$ [black];
$\tau_\pi=\tau_\Pi=\tau_\pi^{\rm (SYM)}$, conformal 
terms in (\ref{eq:evol2}) switched on, $c_\zeta=0$ [gray].
The triangles indicate the largest and smallest times of critical temperature 
corresponding to the last and the first scenarios, respectively.}
\label{fig4}
\end{figure}

Lastly, we study the influence of the relaxation times and the conformal
terms in the evolution.  Figure \ref{fig4} (a) shows $P_z/P$ 
and $P_\perp/P$, using the standard equation of motion (S) for $\Phi$
and $c_\zeta=a_\zeta=1$ using both the estimate for relaxation times $\tau_\pi$ 
and $\tau_\Pi$ from kinetic theory (kin), and the lower estimate (SYM). We also 
show the lower $\tau_\pi$ (kin) with the additional conformal terms (C) 
switched on and $c_\zeta=1$, and the same with $c_\zeta=0$. Figure
\ref{fig4} (b) shows the relative contributions to entropy production in these 
four scenarios as before.

Obviously, larger relaxation times lead to increased entropy production and
larger deviations from equilibrium. This is very clear for the shear 
contributions at early times. The effect of different relaxation times
seems to be very much suppressed for the bulk stress. In fact, after 
$\tau \approx 2$ fm/$c$ we again see a universal evolution
of the system for fixed bulk viscosity. As expected the additional conformal
terms in (\ref{eq:evol2}) lead to a suppression of the shear stress which
manifests itself in a rapid relaxation away from the maximum anisotropic 
initial condition. Enforcing conformal symmetry leads to smaller anisotropies
between the transverse and longitudinal direction and the system is generally
closer to equilibrium. This agrees with the recent observation of Song and Heinz 
\cite{Song:2008si} made in the context of a study of two-dimensional  boost 
invariant hydrodynamics including transverse expansion.

\section{Discussion and Summary}

We can now answer some of the questions posed at the beginning of the paper.
First, independent of the initial conditions we find that the longitudinal
pressure reaches at most half of the equilibrium pressure throughout
the entire lifetime of the quark-gluon plasma phase if the bulk viscosity
is close to the values suggested by lattice QCD ($c_\zeta \approx 1$), even
if the shear viscosity takes its minimum value $\eta_{\mathrm{KSS}}$.
This keeps the system away from equilibrium and reduces the work done in 
the longitudinal expansion. Bulk stress is the main contribution to this 
effect after about 2 fm/$c$ and we find that the evolution of the system
after this time, for a given equation of state at equilibrium, is solely 
determined by the value of the bulk viscosity and
largely independent of the initial conditions and relaxation times.
Scaling our lattice inspired bulk viscosity with factors $c_\zeta > 1$ leads 
to unacceptably small or even negative longitudinal pressure which would
indicate a breakdown of the hydrodynamic picture. These results remain
qualitatively unchanged if the width of the peak in the kinematic bulk viscosity
is varied by a factor of two.

We also found that isotropization of the stress tensor proceeds rather slowly 
with $|P_T - P_z|/P$ as large as 20\% around $T_c$. It was recently pointed
out by Martinez and Strickland \cite{Martinez:2008di} that the anisotropy of
the stress tensor of the quark-gluon plasma may be observable via changes
in the dilepton yield. On the other hand, the strong reduction in the relative 
longitudinal pressure, which may even lead to negative values of $P_z$ 
near $T_c$, is reminiscent of a first-order phase transition, where the 
negative pressure is avoided by the formation of a mixed phase. Under 
favorable conditions, the delay of the expansion caused by mixed phase 
formation can be observed as a directional dependence of the 
identical-particle correlation function in density interferometry
\cite{Pratt:1986cc,Bertsch:1988db,Bertsch:1989vn}. The scenario found
here may show similar effects, but differs from the traditional one by the 
anisotropy of the stress tensor caused by the continued presence of the 
shear viscosity. A realistic exploration of the influence of the bulk viscosity 
on identical particle correlations will the require the hydrodynamical treatment 
of the transverse expansion including both, bulk and shear viscosity.

On the other hand, we find that bulk stress has a rather modest impact on 
entropy production, contrary to some previous expectations. For the
parametrization $\zeta(T)$ considered here (see Fig.\ \ref{fig1}, $c_\zeta=1$) 
the entropy increase due to the bulk viscosity is at most 10\% of the final
entropy. The reason is that the velocity 
gradients due to the longitudinal expansion have decreased significantly 
when the fireball approaches $T_c$. This can also be easily seen by noting 
that we have only shown bulk and shear stress \emph{relative} to the 
equilibrium pressure. While those are comparable, by the time the maximum in 
$-\Pi/P$ around $T_c$ is reached the pressure $P$ has dropped significantly 
and the \emph{absolute} values of $\Pi$ are much smaller than the values 
of $\Phi$ reached at times smaller than 1 fm/$c$.
A bulk viscosity much larger than indicated by our extrapolation of the 
existing lattice QCD results (represented here by scale factors 
$c_\zeta \gg 1$) would be necessary to dominate entropy production after 
decoherence. This statement is independent of the initial condition for 
the bulk stress.

Non-linear terms in the evolution of the shear stress, dictated by 
conformal symmetry, suppress the shear stress and lead to
reduced anisotropies and entropy production. However, these effects are
not large enough to affect the conclusions drawn above qualitatively.

To summarize, large bulk viscosities around $T_c$ lead to prolonged
deviations from equilibrium that could be sizable throughout the entire
lifetime of the quark gluon plasma. Bulk viscosities just slightly larger
than currently favored could easily lead to a breakdown of the 
hydrodynamic approximation around $T_c$. The decreased pressure should 
slow down the expansion of the system and increase the time spent in the 
vicinity of the phase transition.
However, the amount of entropy produced through bulk stress around $T_c$
is smaller than that produced by shear stress at earlier stages of the 
evolution and thus does not result in a large increase of the final 
particle multiplicity, unless the bulk viscosity is much larger than that
considered here.

{\it Acknowledgments:} 
We thank the members of the Yukawa Institute of Theoretical Physics
for their hospitality during the YIPQS Molecule {\sl Entropy Production
Before QGP}. RJF would like to thank P.\ Huovinen for helpful discussions.
This work was supported in part by the U.~S.~Department of Energy 
(grants DE-FG02-05ER41367, DE-AC02-98CH10886), the RIKEN-BNL
Research Center, the Texas A\&M College of Science, and the 
Bundesministerium f\"ur Bildung und Forschung.

\end{document}